\documentclass[lettersize,journal]{IEEEtran}
\usepackage{amsmath,amsfonts}
\usepackage{algorithmic}
\usepackage{algorithm}
\usepackage{array}
\usepackage{color}
\usepackage{xcolor}
\usepackage[caption=false,font=footnotesize,labelfont=rm,textfont=rm]{subfig}
\usepackage{textcomp}
\usepackage{stfloats}
\usepackage{url}
\usepackage{verbatim}
\usepackage{graphicx}
\usepackage{cite}
\usepackage{multirow}
\usepackage{booktabs}
\usepackage{enumerate}
\begin{document}

\title{NaturalL2S: End-to-End High-quality \\Multispeaker Lip-to-Speech Synthesis \\with Differential Digital Signal Processing}

\author{
        Yifan Liang, Fangkun Liu, Andong Li, Xiaodong Li, Chengshi Zheng,~\IEEEmembership{Senior Member,~IEEE}
\thanks{This paper was produced by the IEEE Publication Technology Group. They are in Piscataway, NJ.}
\thanks{Manuscript received April 19, 2021; revised August 16, 2021.}}

\markboth{Journal of \LaTeX\ Class Files,~Vol.~14, No.~8, August~2021}%
{Shell \MakeLowercase{\textit{et al.}}: A Sample Article Using IEEEtran.cls for IEEE Journals}


\maketitle

\begin{abstract}
Recent advancements in visual speech recognition (VSR) have promoted progress in lip-to-speech synthesis, where pre-trained VSR models enhance the intelligibility of synthesized speech by providing valuable semantic information. The success achieved by cascade frameworks, which combine pseudo-VSR with pseudo-text-to-speech (TTS) or implicitly utilize the transcribed text, highlights the benefits of leveraging VSR models. However, these methods typically rely on mel-spectrograms as an intermediate representation, which may introduce a key bottleneck: the domain gap between synthetic mel-spectrograms, generated from inherently error-prone lip-to-speech mappings, and real mel-spectrograms used to train vocoders. This mismatch inevitably degrades synthesis quality. To bridge this gap, we propose Natural Lip-to-Speech (NaturalL2S), an end-to-end framework integrating acoustic inductive biases with differentiable speech generation components. Specifically, we introduce a fundamental frequency (F0) predictor to capture prosodic variations in synthesized speech. The predicted F0 then drives a Differentiable Digital Signal Processing (DDSP) synthesizer to generate a coarse signal which serves as prior information for subsequent speech synthesis. Additionally, instead of relying on a reference speaker embedding as an auxiliary input, our approach achieves satisfactory performance on speaker similarity without explicitly modelling speaker characteristics.
Both objective and subjective evaluation results demonstrate that NaturalL2S can effectively enhance the quality of the synthesized speech when compared to state-of-the-art methods. Our demonstration page is accessible at https://yifan-liang.github.io/NaturalL2S/.
\end{abstract}


\section{Introduction}
\IEEEPARstart{L}{ip}-to-speech synthesis aims to reconstruct speech from silent videos by analyzing lip movements \cite{milner2015}, providing a viable solution for high-quality speech capture in low signal-to-noise ratio environments and remote monitoring applications. However, this task remains exceptionally challenging due to inherent visual ambiguities in phoneme representation. Specifically, most English phonemes exhibit visually confusable articulations (e.g., /p/ vs. /b/), while other phonemes lack distinctive mouth movements (e.g., glottal stops). These limitations explain why the lip-reading accuracy of young adults with normal hearing is only approximately 12\% and rarely exceeds 30\% \cite{altieri2011}. While deep learning technology has driven breakthroughs in Visual Speech Recognition (VSR) with Word Error Rates (WER) below 15\% \cite{ma2023-autoavsr}, progress in lip-to-speech synthesis has been comparatively slower. This disparity is probably attributed to the additional complexities inherent involved in the lip-to-speech task, such as ensuring audio-visual synchronization, capturing speaker-specific vocal characteristics (timbre), and modeling the relationship between prosody and mouth movements. 

Pioneer lip-to-speech studies have focused on either constrained word datasets within controlled environments, such as GRID\cite{cooke2006audio}, TCD-TIMIT\cite{harte2015tcd}, and OuluVS\cite{anina2015ouluvs2}, or open vocabulary datasets from real-world scenarios with specific speakers, such as Lip2Wav\cite{prajwal2020learning}. Many of these methods\cite{milner2015,lecornu2017,ephrat2017,kumar2019,michelsanti2020-vocoderbased} estimate speech parameters from silent video inputs and synthesize speech using statistical parametric vocoders, whereas others\cite{qu2019, prajwal2020learning, hong2021, kim2021, kim2021a, yadav2021, oneata2021, hong2022a, qu2022, wang2022a, varshney2022, he2022flow, sheng2023, kim2023-liptospeech, wang2022, schoburgcarrillodemira2022, niu2023, kefalas2024} directly predict mel-spectrograms, which are then converted into speech waveforms using pre-trained neural vocoders. These studies have undoubtedly advanced the field by establishing fundamental methodologies. However, their practical applicability remains constrained by several limitations. A major limitation is the generalization capability of these models, as most approaches require speaker-specific training data and struggle to synthesize intelligible speech for unseen speakers. Moreover, generating intelligible and natural speech from unconstrained vocabulary remains an ongoing challenge, given the inherent ambiguity of lip movements.

Recent lip-to-speech studies have turned towards more complex scenarios involving arbitrary speakers and unconstrained vocabulary datasets, such as LRW \cite{Chung16}, LRS2 \cite{Chung17}, and LRS3 \cite{afouras2018lrs3}. These datasets introduce substantial challenges, including multi-speaker variability and larger vocabulary sizes. Estimating speaker timbre from silent video remains a significant challenge, and limited visual cues exacerbate the one-to-many mapping problem between visual features and speech. To address speaker identity preservation in multi-speaker lip-to-speech synthesis, some methods incorporate speaker embeddings extracted from raw speech clips using pre-trained speaker encoders \cite{choi2023-intelligible, schoburgcarrillodemira2022, hegde2022, hegde2023, kim2023-liptospeech, prajwal2020learning}. However, obtaining speaker embeddings for unseen individuals during inference remains a fundamental challenge. A vision-guided speaker embedding extractor \cite{choi2023} has been proposed to predict speaker characteristics from visual inputs, mitigating the need for explicit reference speech.

Thanks to advancements in lip-reading, recent lip-to-speech techniques\cite{choi2023-intelligible, choi2023, hsu2023, hegde2023, yemini2023lipvoicer} have demonstrated promising intelligible results for arbitrary speakers in unconstrained settings, effectively mitigating the challenges posed by unconstrained vocabulary. Most of these techniques utilize a pre-trained lip-reading model to extract content features from lip movements. Experimental results have demonstrated that these methods can produce more intelligible speech by introducing language knowledge. For instance, \cite{yemini2023lipvoicer} exploited 
a pre-trained VSR model to predict text for subsequent speech generation, essentially converting lip-to-speech synthesis into a visually-driven text-to-speech (TTS) system. While utilizing a pre-trained state-of-the-art VSR model\cite{ma2023-autoavsr} remarkably reduces the WER of the produced speech, it lacks practicality in scenarios that the text modality data is absent. Alternatively, some researchers\cite{choi2023-intelligible, choi2023, hsu2023} utilized the pre-trained self-supervised learning (SSL) model, Audio-Visual HuBERT (AV-HuBERT)\cite{shi2022learning}, as the original visual front-end. Most of these methods can be viewed as cascaded frameworks that couples pseudo-VSR and pseudo-TTS modules. However, these two components are usually separately trained, which introduces two bottlenecks: 1) This paradigm creates a distributional discrepancy between pseudo-VSR outputs and the ground-truth data expected by downstream pseudo-TTS modules. Specifically, the inherent limitations in pre-trained AV-HuBERT (e.g., lip-reading ambiguity) lead to error propagation and the erroneous semantic predictions from pseudo-VSR will contaminate the pseudo-TTS input space. 2) The multi-stage pipeline inevitably suffers from information degradation, where subtle phonetic details are progressively attenuated through successive processing stages. This cascaded error amplification ultimately manifests as audible artifacts in synthesized speech, leading to diminished speech quality.

To remedy the above-mentioned limitations, this paper proposes \textbf{NaturalL2S}, an innovative end-to-end framework for multi-speaker lip-to-speech task, which leverages a differentiable digital signal processing (DDSP) synthesizer \cite{engel2019} to achieve high-quality and interpretable lip-to-speech synthesis. Inspired by recent lip-to-speech models\cite{choi2023,choi2023-intelligible}, we employ a frozen AV-HuBERT model for video features extraction and visual embeddings generation. Subsequently, a pitch generator and a content generator, comprised of Feed-Forward Transformers (FFT) \cite{ren2020fastspeech}, are used to predict content embeddings and F0 for subsequent corase DSP signal generation. The content generator processes visual embeddings under speech unit supervision to produce content embeddings. Meanwhile, the pitch generator produces F0 through a multi-stage frequency prediction pipeline. Specifically, the video features are upsampled and processed through FFT blocks, along with the content embeddings, to estimate coarse F0. The initial predicted F0 and content embeddings are then used to guide the auxiliary coarse mel-spectrograms prediction. An F0 post predictor utilizes these mel-spectrograms to produce final refined F0 contours. The content embeddings and corrected F0 values drive preliminary coarse DSP waveforms through the DDSP synthesizer. Finally, we utilize a conditional HiFi-GAN to refine the coarse DSP signals into high-quality speech waveforms.

The main contributions of this paper can be summarized as follows:
\begin{enumerate}
	\item This is the first attempt to jointly train the vocoder for the multi-speaker lip-to-speech task to the best of our knowledge. The end-to-end training strategy enables the model to learn more efficient representations and relationships between visual inputs and speech outputs, mitigating the common mismatch issues that existed in previous multi-stage training approaches.
	
	\item  We novely incorporate a DDSP synthesizer for interpretable and high-fidelity speech synthesis. The signals generated by the DDSP module act as acoustic inductive biases, enhancing the generation of harmonic structure and aperiodic components of the synthesized speech. This approach aims to provide additional variance information to simplify the data distribution complexity and enhance synthesized speech quality, as highlighted in \cite{ren2022revisiting}. 
	
	\item Inspired by \cite{choi2023}, NaturalL2S generates speech waveforms without requiring reference speaker embedding, which makes it more practical for real-world scenarios that speaker identity might not be accessed in advance. Evaluations demonstrate that our model remarkably achieves high speaker similarity even without explicit speaker embedding supervision during training.
\end{enumerate}

The remainder of this paper is organized as follows. Section~\uppercase\expandafter{\romannumeral2} reviews previous related work. Section~\uppercase\expandafter{\romannumeral3} describes the details of our proposed lip-to-speech method. Section~\uppercase\expandafter{\romannumeral4} and Section~\uppercase\expandafter{\romannumeral5} give descriptions of the experimental settings and results, respectively. In Section~\uppercase\expandafter{\romannumeral6}, the conclusion is presented.

\section{Related Work}
\subsection{Multi-speaker Lip-to-Speech Synthesis}
Compared to the prior datasets focusing on constrained experimental settings, lip-to-speech research based on multi-speaker datasets has shown enhanced scalability and practicality for real-world applications. Scalable video-to-speech synthesis (SVTS) \cite{schoburgcarrillodemira2022} is the first attempt to produce intelligible speech from silent videos on an unconstrained multi-speaker dataset. This method incorporates a scalable mel-spectrogram generator and a pre-trained neural vocoder. The generator includes a 3D Convolution and ResNet-18\cite{he2016deep} encoder followed by a conformer\cite{gulati2020conformer} decoder. The width and depth of the Conformer blocks are adjusted based on the scale of datasets in the experiments. At the inference phase, SVTS exploits the pre-trained Parallel WaveGAN\cite{yamamoto2020parallel} as the vocoder to transform the generated mel-spectrogram into the waveform. Lip-to-speech Multi-Task\cite{kim2023-liptospeech} further improves this framework by introducing the multi-task learning strategy to simultaneously supervise the text and audio modality. Specifically, unlike previous works, which solely use the reconstruction loss to supervise acoustic features, both the feature level and output-level content supervision are also added to guide the model toward generating more intelligible speech. However, it employs the Griffin-Lim algorithm as the vocoder, so the quality of synthesized speech was limited. Hedge et al. \cite{hegde2022} leveraged the probabilistic nature of a variational autoencoder (VAE) to mitigate the one-to-many mapping challenge inherent in lip-to-speech synthesis. By explicitly modeling speech content uncertainty through latent space variational inference, their VAE-GAN framework captures the inherent variability between lip movements and corresponding speech, avoiding deterministic overfitting in one-to-many mapping problems.

Several works that adopt pre-trained models to obtain text from silent videos have shown considerable performance superiority. Hedge et al. \cite{hegde2023} employed a pre-trained lip-reading model presented in \cite{prajwal2022} to acquire text predictions and visual representations of each video frame. Additionally, the scaled dot-product attention module\cite{vaswani2017attention} is exploited to learn the correspondence between the text phoneme features and visual representations, yielding the content representations. All representations are finally passed through a cascade pipeline consisting of a spectrogram decoder and a pre-trained BigVGAN\cite{lee2023} to generate the final speech waveforms. LipVoicer\cite{yemini2023lipvoicer} also utilizes a pre-trained lip-reading model\cite{ma2023-autoavsr} to infer the text from the silent videos, which helps distinguish the ambiguous syllables visually. The predicted text is used to provide classifier guidance\cite{ho2022classifier} to improve the performance of a diffusion-based mel-spectrogram generator\cite{kong2020diffwave}. As a consequence of the utilization of a powerful lip-reading model, LipVoicer exhibits exceptionally high intelligibility in generated speech samples. However, the performance of these approaches heavily relies on the efficacy of the pre-trained lip-reading model, and methods without text modality dependence are more practical in real-world environments, in which not all videos have corresponding text transcripts. 

The emergence of large-scale audio-visual self-supervised models such as AV-HuBERT \cite{shi2022learning} provides a text-free paradigm for lip-to-speech synthesis through cross-modal representation learning. Pre-trained on massive audio-visual corpora (LRS3 and VoxCeleb2 \cite{chung2018voxceleb2}), AV-HuBERT employs a masked prediction objective where the model reconstructs discrete cluster labels for masked audio or visual segments. An iterative offline k-means process refines these cluster assignments during training. This multi-modal pretraining strategy enables AV-HuBERT to learn transferable representations, which generalize effectively across diverse downstream tasks (e.g., lip-reading, lip-to-speech synthesis). ReVISE \cite{hsu2023} demonstrates this capability by employing HuBERT \cite{hsu2021hubert} to generate speech units that connect two subsystems: 1) a pseudo audio-visual speech recognizer (P-AVSR) initialized with AV-HuBERT, and 2) a unit-driven HiFi-GAN synthesizer (P-TTS). The framework trains both components using targets from a fine-tuned HuBERT-BASE, achieving superior intelligibility over non-AV-HuBERT baselines. In \cite{choi2023-intelligible}, researchers utilized the pre-trained AV-HuBERT to generate both mel-spectrogram and speech units as the input of HiFi-GAN. However, they observed that the intermediate mel-spectrograms suffered from blurriness and noise artifacts compared to the ground truth. This can be attributed to the inherent limitations of input information from silent video. To mitigate the mismatch between the generated mel-spectrograms and real mel-spectrograms, the authors proposed an augmentation technique to simulate the blurry and noisy generated mel-spectrogram during multi-input vocoder training.

Focusing on the difficulty of getting speaker timbre, most of the above-mentioned methods provide a prior speaker embedding to assist the model during both training and inference stages since timbre information is difficult to obtain from videos accurately. However, relying on speaker embedding limits the application scope for real-world scenarios. To solve this problem, DiffV2S\cite{choi2023} addresses this by introducing a vision-guided speaker embedding extractor to predict speaker characteristics. The speaker characteristics are regarded as the conditions of the diffusion model\cite{sohl2015deep} to generate a mel-spectrogram, which is then converted to a waveform using HiFi-GAN. 

\subsection{Differentiable Digital Signal Processing}
Differentiable digital signal processing (DDSP) \cite{engel2019} combines traditional digital signal processing with deep learning to employ the inherent acoustic knowledge in classic signal processing components like filters or oscillators. This approach enables interpretable high-fidelity audio generation. Originally developed for interpretable high-fidelity audio generation, the DDSP model comprises an encoder that maps the input log mel-spectrogram to a latent representation and a decoder that predicts parameters for additive and filtered noise synthesizers.

The modularity and interpretability of DDSP make it highly effective for various audio synthesis tasks, including singing voice synthesis \cite{zhang2022}, artificial reverberation\cite{lee2022differentiable}, and vocoders\cite{li2023}. Our work is inspired by \cite{zhang2022}, in which researchers designed a conditional HiFi-GAN framework to enhance an end-to-end variational autoencoder singing voice synthesis system. This framework integrates a DDSP synthesizer to produce periodic and aperiodic signals from the latent representation, which are then used as an auxiliary input for the multi-receptive field fusion (MRF) module in HiFi-GAN.

By incorporating a DDSP synthesizer, NaturalL2S leverages these advantages to not only achieve high-quality speech but also surpass many state-of-the-art lip-to-speech models in both fidelity and naturalness.

\section{Model Architecture}

\begin{figure*}[htbp] 
	\centering
	\includegraphics[width=6.5in]{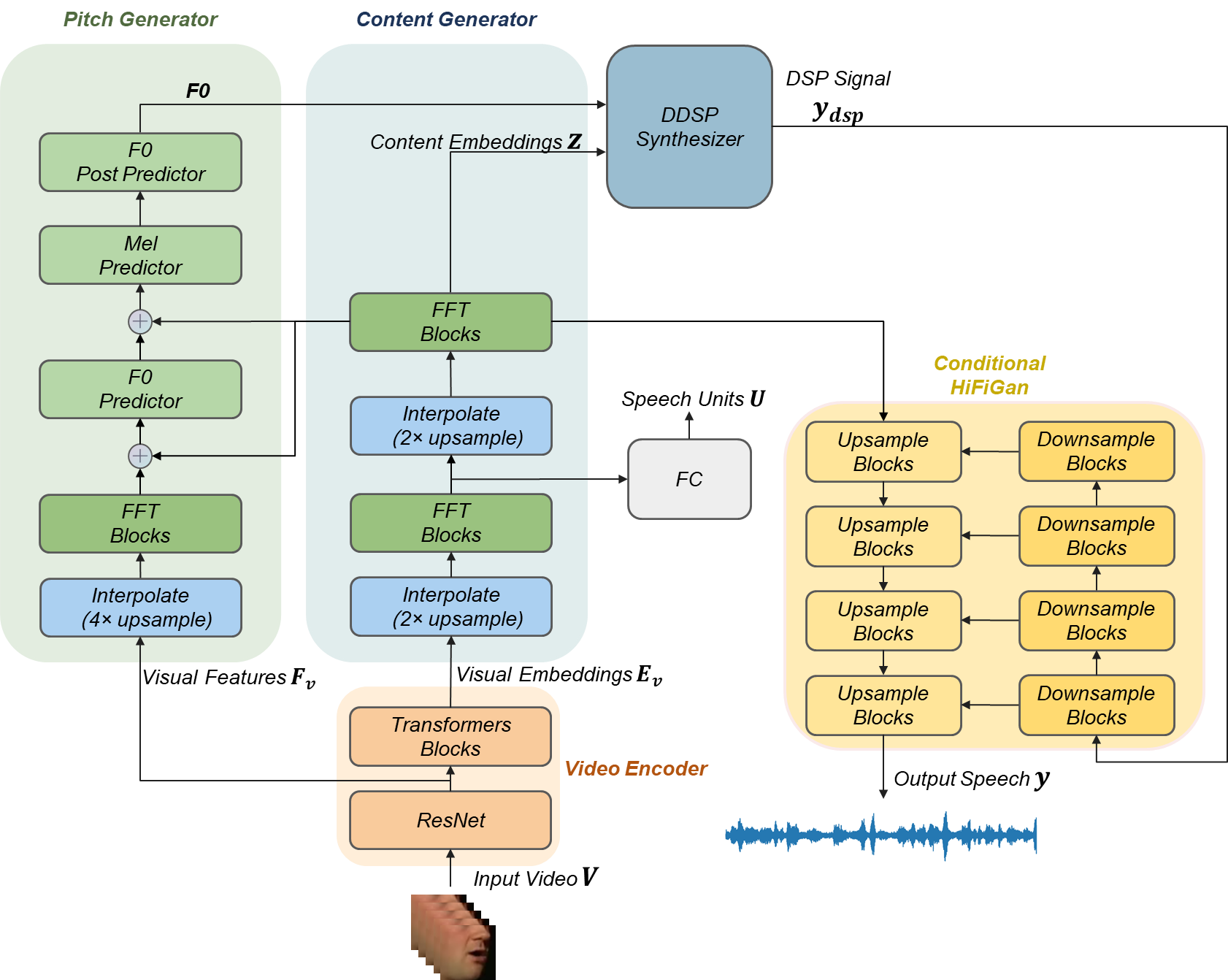} 
	\caption{\centering Overview of the proposed NaturalL2S.}
	\label{fig:figure1}
\end{figure*}

In this paper, we denote the silent video containing lip movements as \( \mathbf{X}=\{ \mathbf{x}_{1},\ldots,\mathbf{x}_{T}\}\in\mathbb{R}^{T\times H\times W\times C} \), where \( T \) represents the number of the input video frames, and \( H \), \( W \), and \( C \) represent the height, width, and number of channels for each video frame, respectively. The corresponding mel-spectrograms and speech units are, respectively, denoted as \( \mathbf{Y}=\{ \mathbf{y}_{1},\ldots,\mathbf{y}_{M}\}\in\mathbb{R}^{M\times N} \) and \( \mathbf{U}=\{ u_{1},\ldots,u_{L}\}\in\mathbb{R}^{L} \). Here, \( M \) represents the number of frames in the mel-spectrogram, \( N \) represents the number of mel filter banks, and \( L \) represents the length of the speech unit sequence. In our experiments, the \( L\) is set to \(2T\) while the \(M\) is set to \(4T\).

The NaturalL2S model architecture, depicted in Fig. \ref{fig:figure1}, encompasses five key components: a pre-trained video encoder, a content generator, a pitch generator, a DDSP synthesizer, and a conditional HiFi-GAN. Initially, we leverage the pre-trained AV-HuBERT model to extract visual embeddings \(\mathbf{E}_{v}\) from silent video. These visual embeddings are then passed through the content generator and the pitch encoder, which predict the fundamental frequency (F0) and generate the content embedding \(\mathbf{Z}\), respectively. Both outputs are subsequently fed into the DDSP synthesizer. The DDSP synthesizer contains two main components: a harmonic synthesizer and a noise synthesizer. The harmonic synthesizer generates the periodic components of the speech signal, while the noise synthesizer handles the aperiodic components. The coarse DSP signals generated by the DDSP synthesizer serve as a conditioning factor for the following vocoder stage. Finally, the neural vocoder HiFi-GAN takes the coarse DSP signals and content features as inputs and generates the final speech waveform. Each component of the architecture is discussed in detail in the following sections.

\subsection{Content Generator}
The content generator utilizes a frozen AV-HuBERT model as a fixed video encoder to extract visual embeddings \( \mathbf{E}_{v} \), avoiding the computational overhead of full model fine-tuning. Visual embeddings \( \mathbf{E}_{v} \)  are extracted from the output of the transformer component in AV-HuBERT, which can be described as:
\begin{equation}
  \label{deqn_ex1a}
  \mathbf{E}_{v} = VideoEncoder(\mathbf{X})\in\mathbb{R}^{T\times D_{e}},
\end{equation}
where \( D_{e} \) represents the embedding dimension. These embeddings encode phonetic and linguistic information critical for speech reconstruction.

Motivated by \cite{choi2023-intelligible}, we incorporate quantized Self-Supervised Learning (SSL) speech units from a pre-trained SSL model \cite{hsu2021hubert} to provide robust linguistic supervision. Specifically, the visual embeddings \(\mathbf{E}_{v}\)  are interpolated to match the target speech unit sampling rate and processed through several FFT blocks and a fully connected (FC) layer to generate the speech units \(\widehat{\mathbf{U}}\). This operation can be described as:
\begin{equation}
  \label{deqn_ex1a}
  \widehat{\mathbf{U}} = \mathcal{F}(FFT(\mathcal{I}(\mathbf{E}_{v}))) \in \mathbb{R}^{L},
\end{equation} 
where \( \mathcal{F} \) represents the FC layer and the interpolation operation is denoted as \( \mathcal{I} \). During training, the label-smoothed cross-entropy loss is adopted to optimize the model:
\begin{equation}
  \label{deqn_ex1a}
  \mathcal{L}_{ce} = -\sum_{t=1}^{L} \sum_{c=1}^{C} \left[(1 - \epsilon) \mathbf{U}_{t,c} + \frac{\epsilon}{C}\right] \log p(\widehat{\mathbf{U}}_{t,c}),
\end{equation}
where \(C\) is the cluster number of speech units and \(\epsilon\) is the the label smoothing factor. The intermediate features are further upsampled and processed through additional FFT blocks to produce the final content embeddings:
\begin{equation}
  \label{deqn_ex1a}
  \mathbf{Z} = ContentEncoder(\mathbf{E}_{v})\in\mathbb{R}^{M\times D_{m}},
\end{equation}
where \( D_{m} \) is the hidden dimension of the model. These embeddings serve as crucial content representation for speech synthesis and as auxiliary inputs into the pitch generator for generating coarse F0 and mel-spectrogram.

\subsection{Pitch Generator}
The pitch generator module predicts F0 trajectories from visual features to ensure natural prosodic variations in synthesized speech, which are crucial for the expressivity and naturalness of the synthesized speech. Recent findings \cite{choi2023} demonstrate that visual features from the visual front-end of the pre-trained AV-HuBERT can effectively capture speaker-specific acoustic characteristics through prompt tuning. This suggests that these visual features inherently contain acoustic information for prosodic feature extraction. Inspired by this work, the pitch generator jointly processes both the visual features \( \mathbf{F}_{v} \) and the content embeddings \(\mathbf{Z}\) for robust F0 prediction. We employ a cascade of FFT blocks to progressively extract prosodic information. Specifically, we utilize initial FFT blocks to process unsampled features and dedicated FFT blocks as the F0 predictor, the mel predictor, and the F0 post predictor.

The technical implementation starts by upsampling the visual features \( \mathbf{F}_{v} \) fourfold to match the mel-spectrogram sampling rate. These upsampled visual features are passed through initial FFT blocks and then combined with content embedding \(\mathbf{Z}\) to predict coarse log-F0 (\(\widehat{\mathbf{lf0}}\)):
\begin{equation}
\label{deqn_ex1a}
\widehat{\mathbf{lf0}} = F0Predictor(FFT(\mathcal{I}(\mathbf{F}_{v}))+\mathbf{Z})\in\mathbb {R}^{M}.
\end{equation}

Subsequently, the coarse \(\widehat{\mathbf{lf0}}\) and content embeddings \(\mathbf{Z}\) jointly predict an intermediate mel-spectrogram:
\begin{equation}
	\label{deqn_ex1a}
	\widehat{\mathbf{Y}}_{pred} = MelPredictor(\widehat{\mathbf{lf0}}+\mathbf{Z})\in\mathbb{R}^{M\times D_{m}}.
\end{equation} 

The auxiliary mel-spectrogram \(\widehat{\mathbf{Y}}_{pred} \) guides the F0 post predictor in generating final refined F0 prediction:
\begin{equation}
	\label{deqn_ex1a}
	\widehat{\mathbf{lf0}}_{post} = F0PostPredictor(\widehat{\mathbf{Y}}_{pred})\in\mathbb{R}^{M}.
\end{equation}

Due to the limited variance information from lip movements, the generated mel-spectrogram tends to be coarse and overly smooth. Consequently, it is not directly utilized in the speech synthesis process, as this could potentially impair vocoder training.

During training, we apply L1 loss to supervise the generation of the predicted  \(\widehat{\mathbf{lf0}}\), the auxiliary mel-spectrogram \(\widehat{\mathbf{Y}}_{pred} \) and the refined \(\widehat{\mathbf{lf0}}_{post}\) as follows:
\begin{equation}
  \label{deqn_ex1a}
  \mathcal{L}_{F0} = \Vert \mathbf{lf0}-\widehat{\mathbf{lf0}} \Vert_1 + \Vert \mathbf{lf0}-\widehat{\mathbf{lf0}}_{post}\Vert_1,
\end{equation}
\begin{equation}
  \label{deqn_ex1a}
  \mathcal{L}_{mel} = \Vert \mathbf{Y}-\widehat{\mathbf{Y}}_{pred} \Vert_1,
\end{equation} 
where \(\mathbf{lf0}\) and \(\mathbf{Y}\) represent the ground truth of the log F0 and that of the mel-spectrogram, respectively.
\begin{figure*}[ht] 
	\centering
	\includegraphics[width=5.8in]{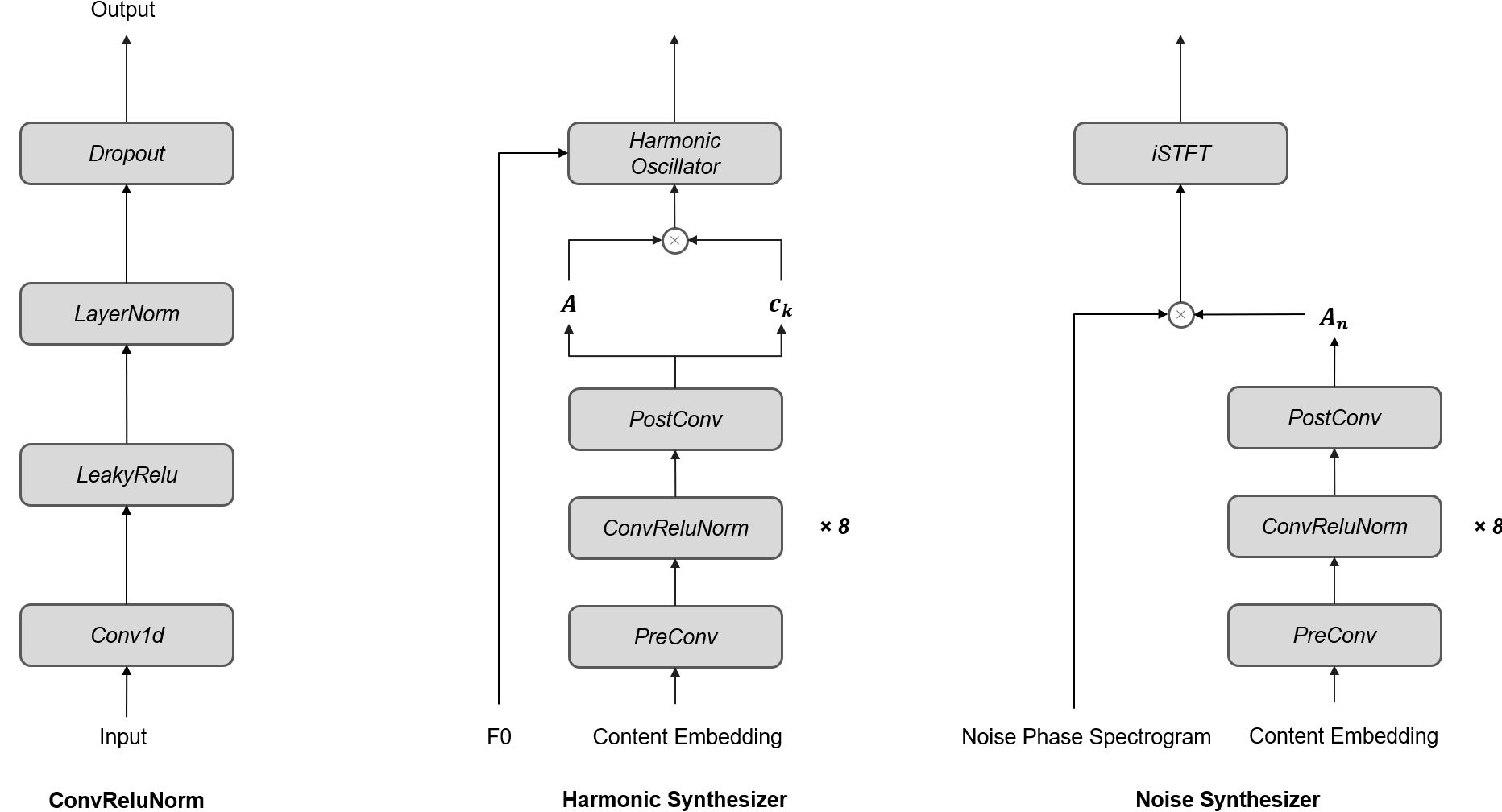} 
	\caption{\centering The internal operation of the DDSP Synthesizer. }
	\label{fig:figure2}
\end{figure*}

\subsection{DDSP Synthesizer}
The DDSP synthesizer module is a core component in NaturalL2S, which synthesizes the initial conditional coarse DSP signal for subsequent waveform generation. According to the harmonic-plus-noise model of the speech\cite{stylianou1996harmonic}, the coarse DSP signal \(\mathbf{y}_{dsp}(t)\) generated by DDSP can be decomposed into a periodically harmonic part \(\mathbf{y}_{h}(t)\) and an aperiodic part \(\mathbf{y}_{n}(t)\), it can be represented as:
\begin{equation}
  \label{deqn_ex1a}
  \mathbf{y}_{dsp}(t) = \mathbf{y}_{h}(t) + \mathbf{y}_{n}(t).
\end{equation}

We adopt the harmonic and noise synthesizer of DDSP\cite{engel2019} to generate coarse signals from the content embedding \(\mathbf{Z}\) and predicted \(\widehat{\mathbf{lf0}}_{post}\). As shown in the middle part of Fig. \ref{fig:figure2}, the harmonic synthesizer generates the periodic signals \(\mathbf{y}_{h}(t)\), capturing the harmonic structure essential for vowel sounds and other voiced elements. This process of producing periodic sinusoidal signals can be formulated as follows:
\begin{equation}
  \label{deqn_ex1a}
  \mathbf{y}_{h}(t) =  A(t)\sum_{k=1}^{k}c_{K}(t)sin(\phi_{k}(t)),
\end{equation}
where \(K\) is the number of sinusoidal component, \(A(t)\) is the global amplitude corresponding to the time step \(t\) and \(c_{k}(t)\) is the normalized amplitude of the \(k\)-th harmonic component. The \(\widehat{\mathbf{f0}}_{post}\) sequence is interpolated to the sample level to produce the instantaneous phase \(\phi_{k}(t)\) as follows:
\begin{equation}
  \label{deqn_ex1a} 
  \phi_{k}(t) = 2\pi\sum_{\tau=0}^{t}k\widehat{\mathbf{f0}}_{post}({\tau})+\phi_{0,k},
\end{equation}
where \(\phi_{0,k}\) is randomised with values in \([-\pi,\pi]\). The parameters \(A\), \(c_{k}\) are estimated for each frame by the ConvReLUNorm module, which is shown in the left part of Fig. \ref{fig:figure2}, and upsampled to the time domain through the interpolate operation.

On the other hand, the noise synthesizer takes the content embedding \(\mathbf{Z}\) as input and then produces aperiodic signals \(\mathbf{y}_{n}(t)\) to capture the aperiodic components such as fricatives and plosives, which can be represented as :
\begin{equation}
  \label{deqn_ex1a} 
  \mathbf{y}_{n}(t) = iSTFT(A_{n},P),
\end{equation}
where \(P\) is the phase spectrogram of a uniform noise signal in domain \([-\pi,\pi]\) and the amplitude spectrogram \(A_{n}\) is estimated by the ConvReLUNorm module. The aperiodic components of speech waveforms are obtained through inverse short-time Fourier transform (iSTFT). The entire process is displayed on the right part of Fig. \ref{fig:figure2}. 

To guide the waveform generation of the DDSP synthesizer, we employ L1 loss between the mel-spectrogram of DSP signal \(\mathbf{y}_{dsp}(t)\) and the target speech \(\mathbf{y}(t)\) as follows,
\begin{equation}
  \label{deqn_ex1a} 
  \mathcal{L}_{DSP} = \Vert \mathbf{Y} -\mathbf{Y}_{DSP} \Vert_1,
\end{equation}
where \(\mathbf{Y}_{DSP}\) is denoted as the mel-spectrogram of the generated coarse DSP waveform.

\subsection{Conditional HiFi-GAN}
The GAN-based vocoders can produce realistic outputs through the adversarial training strategy. However, in the lip-to-speech task, the lack of substantial variation information causes the model to fail in learning speech characteristics when relying solely on GANs. The coarse DSP signal generated by our DDSP module contains substantial prior acoustic information, making it an effective choice as a conditioning input. Therefore, we adopt a conditional HiFi-GAN as the vocoder to produce speech, integrating the inductive bias of the DDSP with GAN to generate more natural and expressive speech. Meanwhile, we use the end-to-end training strategy to further improve speech quality. Instead of using mel-spectrogram, we utilize the content embedding \(\mathbf{Z}\) and the DSP signal \(\mathbf{y}_{dsp}\) to generate the waveform. The synthesis process can be expressed as:
\begin{equation}
\label{deqn_ex1a} 
\widehat{\mathbf{y}}(t) = G(\mathbf{Z}, \mathbf{y}_{dsp}),
\end{equation}
where ${{G}}\left(  \cdot  \right)$ represents the operation of the generator module in conditional HiFi-GAN.
Specifically, the coarse DSP signal \(\mathbf{y}_{dsp}\) is used as the conditioning factor to guide the generator to produce high-quality speech waveforms. We progressively downsample the coarse DSP signal through strided 1D convolutions to match the upsample rates in HiFi-GAN. At each upsample block in HiFi-GAN, the intermediate downsampled features are concatenated channel-wise with the upsampled features from the transposed 1D convolutions and then passed through the MRF module, progressively generating the waveform.

Similar to\cite{kong2020}, MultiPeriod Discriminator (MPD) and Multi-Scale Discriminator (MSD) are adopted. During the training stage, we use the least-square loss as the GAN loss and also use the L1 loss as the feature matching loss like \cite{kong2020}. The loss functions are described as:
\begin{equation}
	\label{deqn_ex1a}
	\begin{aligned}
	\mathcal{L}_{adv}(D; G)= &\ \mathbb{E}_{(\mathbf{s},\mathbf{Z}, \mathbf{y}_{dsp})}\Big[(D(\mathbf{s})-1)^2 \\ &+(D(G(\mathbf{Z},\mathbf{y}_{dsp})))^2\Big],
	\end{aligned}
\end{equation} 
\begin{equation}
  \label{deqn_ex1a} 
  \mathcal{L}_{adv}(G; D)=\mathbb{E}_{(\mathbf{Z}, \mathbf{y}_{dsp} )}\Big[(D(G(\mathbf{Z},\mathbf{y}_{dsp}))-1)^2\Big],
\end{equation}
\begin{equation}
\begin{aligned}
  \label{deqn_ex1a} 
  \mathcal{L}_{fm}(G; D)= &\ \mathbb{E}_{(\mathbf{s},\mathbf{Z}, \mathbf{y}_{dsp})}\Big[\sum_{l=1}^{T_{L}}\frac1{N_l}\|D^l(\mathbf{s})\\ & -D^l(G(\mathbf{Z}, \mathbf{y}_{dsp}))\|_1\Big],
 \end{aligned}
\end{equation}
where \({D} (\cdot) \) represents the operation of the discriminator module, ${{\mathcal{L}_{adv}}}$ represents the adversarial loss for training both the generator and the discriminator, and ${{\mathcal{L}_{fm}}}$ represents the feature matching loss. In \({\mathcal{L}_{fm}} (\cdot)\), \(T_{L}\) denotes the total number of layers in the discriminator, \(D^{l}\) and \(N_{l}\) denotes the features and the number of features in the \(l\)-th layer of discriminator, respectively.

Moreover, according to \cite{kong2020}, the L1 loss between the mel-spectrogram of the generated speech and the ground truth mel-spectrogram is used as the reconstruction loss:
\begin{equation}
  \label{deqn_ex1a}
  \mathcal{L}_{recon} = \Vert \mathbf{Y}-\widehat{\mathbf{Y}} \Vert_1,
\end{equation} 
where \(\widehat{\mathbf{Y}}\) is denoted as the mel-spectrogram of the generated speech waveform.

\subsection{Loss Function}
For convenience, we divide our proposed model into two parts: the generation part including the pitch generator, content generator, and DDSP synthesizer, and the vocoder part consisting of the conditional HiFi-GAN. The loss function \(\mathcal{L}_{gen}\) for the generation part can be expressed as:
\begin{equation}
\label{deqn_ex1a}
\begin{aligned}
\mathcal{L}_{gen} = &\ \lambda_{ce} \mathcal{L}_{ce} + \lambda_{DSP} \mathcal{L}_{DSP} + \lambda_{F0} \mathcal{L}_{F0} \\ &+ \lambda_{mel} \mathcal{L}_{mel},
\end{aligned}
\end{equation} 
where ${\lambda_{ce}}$, ${\lambda_{DSP}}$, ${\lambda_{F0}}$ and ${\lambda_{mel}}$ represent the weight of all the loss items in ${{\mathcal{L}_{gen}}}$. Focusing on the vocoder part, the loss function for training the generator and discriminator can be separately described as:
\begin{equation}
\label{deqn_ex1a}
\mathcal{L}_{G} =  \mathcal{L}_{adv}(G; D)+\mathcal{L}_{fm}(G; D)+\lambda_{recon} \mathcal{L}_{recon},
\end{equation} 
\begin{equation}
	\label{deqn_ex1a}
	\mathcal{L}_{D} =  \mathcal{L}_{adv}(D; G),
\end{equation} 
where ${\lambda_{recon}}$ reprensents the weight of ${{\mathcal{L}_{recon}}}$.\par
With the end-to-end training strategy, the total loss function for training our NaturalL2S can be expressed as follows:
\begin{equation}
\label{deqn_ex1a}
\mathcal{L}_{total} = \mathcal{L}_{gen} + \mathcal{L}_{G}.
\end{equation} 

\section{Experiments}
\subsection{Datasets}
We conduct experiments on two publicly available multi-speaker audio-visual datasets, namely LRS2 \cite{Chung17} and LRS3 \cite{afouras2018lrs3}, to validate the effectiveness of introducing acoustic priors as an inductive bias for high-quality speech generation. The LRS2 dataset, collected from BBC television programs, comprises 144,300 short video clips featuring thousands of speakers, with a total duration of approximately 200 hours. The LRS3 dataset, collected from TED and TEDx talks, includes approximately 150,000 short audio clips uttered by over 5000 speakers, with a total duration of around 439 hours. The video frame rate and audio sample rate in both datasets are 25 Hz and 16 kHz, respectively. Both datasets share a video frame rate of 25 Hz and an audio sample rate of 16 kHz. We follow the original data split for training, validation, and testing.

\subsection{Preprocessing}
For video preprocessing, we employ dlib \cite{king2009dlib} to detect facial landmarks and align each frame to a reference mean face. Subsequently, we crop an 88 × 88 mouth region and convert it to grayscale for model input. In audio processing, we reconstruct and predict mel-spectrograms from raw waveforms. This involves applying a mel-filterbank to spectrograms derived by the Short-Time Fourier Transform (STFT). The specific parameters are set to 80 mel-filterbank numbers, 640 FFT size, 640 window size, and 160 hop size. The resulting mel-spectrogram has a sample rate of 100Hz, which is four times the video frame rate.

We utilize a pre-trained SSL model, HuBERT, to extract linguistic features. These features are then quantized using a pre-trained 200-cluster K-means model to obtain target speech units for content supervision, following the implementation of \cite{choi2023-intelligible}. The sample rate of speech units and fundamental frequency is set to 50Hz, which is double the video frame rate. Finally, the RAPT algorithm \cite{talkin1995robust} is used in the pitch generator to extract the target fundamental frequency

\subsection{Model Architecture}
\subsubsection{Video Encoder}
In this work, we utilize the self-training Large AV-HuBERT model fine-tuned for visual speech recognition as our visual front-end. The AV-HuBERT comprises a modified ResNet-18 and a stack of 24 Transformer layers. It produces 1024-dimensional video features from ResNet-18 and visual embeddings from the output of the final Transformer layer.

\subsubsection{Content Generator and Pitch Generator}
The FFT blocks in the pitch and content generators are adapted from FastSpeech2 \cite{ren2020fastspeech}. Each FFT block consists of 6 stacked Transformer blocks, with 2 attention heads, a hidden dimension of 256 and a feed-forward layer dimension of 768. The F0 predictor and mel-predictor each contain a single FFT block. The output dimension of the F0 predictor is projected to 1, while for the mel-predictor, it is projected to 80. The remaining configurations are similar to those of the FFT blocks.

\subsubsection{DDSP Synthesizer}
The DDSP synthesizer, inspired by the architecture detailed in \cite{zhang2022}, consists of a harmonic synthesizer and a noise synthesizer. The harmonic synthesizer estimates the total amplitude and loudness of each harmonic using a network with ConvReLUNorm blocks. These estimated parameters are then used by an additive oscillator to generate harmonic waveforms. Similarly, the noise synthesizer employs ConvReLUNorm blocks to predict the amplitude spectrogram of the aperiodic signal, used for reconstructing the aperiodic parts of speech with the inverse Short-Time Fourier Transform (iSTFT). We set the number of harmonics to 32 and the FFT size to 1024.

\subsubsection{Conditional HiFi-GAN}
The conditional HiFi-GAN combines the strengths of the DDSP signal waveform and the content embedding. The content embedding is upsampled to match the dimensionality of the downsampled DDSP signal. Specifically, we upsample the content embedding [5,4,4,2] times with upsample blocks and downsample the DDSP signal [2,4,4,5] times with downsample blocks accordingly. In essence, the upsample and downsample blocks ensure compatible dimensions for combining information from the DDSP signal and the content embedding within the HiFi-GAN architecture. The hidden dimension, kernel size in the MRF module, kernel size in the transposed convolutions, and dilation rates in MRF are set to 512, [3,7,11], [11,8,4,4], and [[1,3,5], [1,3,5], [1,3,5]] respectively. For the settings of the Multi-Period Discriminator (MPD) and Multi-Scale Discriminator (MSD), we follow the configurations in \cite{kong2020}.

\begin{table*}[ht]
	\setlength{\abovecaptionskip}{0pt}
	\centering
	\footnotesize
	\caption{Objective evaluation results of different models on LRS2 dataset. $^*$ denotes metric inferences with reference speaker embedding} \label{tbl:1}
	\begin{tabular}{lcccccccc}
		\toprule
		\textbf{Model}  & STOI-Net ↑ & DNS-MOS ↑ & WER ↓ & LSE-C ↑ & LSE-D ↓ & F0pcc ↑ &  Voice Similarity ↑ \\
		\midrule
		VCA-GAN  & 0.51 & 2.25 & 100.3\% & 3.35 & 10.75 & 0.06 & 0.44 \\
		Multi-task  & 0.71 & 2.37 & 50.9\% & 7.00 & 7.17 & 0.38 & 0.55$^*$ \\
		DiffV2S  & 0.89 & 2.92 & 48.9\% & 7.39 & 6.93 & 0.54 & 0.58 \\
		Lip2Speech Unit & 0.86 & 2.67 & \textbf{33.5\%} & 8.45 & 6.00 & 0.58 & \textbf{0.72}$^*$ \\
		\textbf{Proposed} & \textbf{0.92} & \textbf{2.95} & 36.5\% & \textbf{8.74} & \textbf{5.84} & \textbf{0.61} & 0.63 \\
		Ground Truth & 0.91 & 3.01 & 1.35\% & 8.25 & 6.26 & -- & -- \\
		\bottomrule
	\end{tabular}
\end{table*}

\begin{table*}[!h]
	\setlength{\abovecaptionskip}{0pt}
	\centering
	\footnotesize
	\caption{Objective evaluation results of different models on LRS3 dataset. $^*$ denotes metric inferences with reference speaker embedding} \label{tbl:2}
	\begin{tabular}{lcccccccc}
		\toprule
		\textbf{Model}  & STOI-Net ↑ & DNS-MOS ↑ & WER ↓ & LSE-C ↑ & LSE-D ↓ & F0pcc ↑ & Voice Similarity ↑  \\
		\midrule
		VCA-GAN  & 0.64 & 2.27 & 89.6\% & 5.23 & 8.90 & 0.13 & 0.42 \\
		Multi-task  & 0.68 & 2.36 & 56.8\% & 5.20 & 8.82 & 0.37 & 0.45$^*$ \\
		DiffV2S  & 0.93 & 3.22 & 35.8\% & 7.21 & 7.30 & 0.55  & 0.63 \\
		Lip2Speech Unit & 0.90 & 2.89 & \textbf{27.7\%} & 7.94 & 6.59 & 0.58 & \textbf{0.76}$^*$ \\
		\textbf{Proposed} & \textbf{0.94} & \textbf{3.23} & 30.4\% & \textbf{8.08} & \textbf{6.46} & \textbf{0.65} & 0.63 \\
		Ground Truth & 0.93 & 3.20 & 0.71\% & 7.63 & 6.89 & -- & -- \\
		\bottomrule
	\end{tabular}
\end{table*}

\subsection{Training}
We train the NaturalL2S model using the AdamW optimizer \cite{loshchilov2017decoupled} with \(\beta_{1}=0.8\), \(\beta_{1}=0.99\) and weight decay \(\lambda= 0.01\). The ${\lambda_{mel}}$, ${\lambda_{ce}}$, ${\lambda_{F0}}$, ${\lambda_{DSP}}$ and ${\lambda_{recon}}$ weight in loss function is separately set to 1, 20, 20, 45, 45. The learning rate is initially set to \(1 \times10^{-4}\) and managed by an exponential decay curve. However, training all networks end-to-end is computationally intensive and inefficient. To mitigate GPU memory usage and reduce training time, we adopt a training strategy inspired by \cite{kim2021conditional}, where we randomly sample slices from the content embedding and DDSP signal to generate waveforms. Simultaneously, we extract corresponding slices from the ground truth speech as training targets. During the training of the DDSP synthesizer, ground-truth F0 values are used for harmonic signal generation, while predicted F0 values are used for waveform generation during inference. The waveform sample clip is set to 5,920, corresponding to content embedding slices with a length of 37. It takes 500k steps for training until convergence for NaturalL2S on each dataset when trained on a single NVIDIA A100 with a batch size of 48. We employ mixed precision training to accelerate the training process while maintaining comparable accuracy.

\subsection{Evaluation Metrics}
Following \cite{yemini2023lipvoicer}, we adopt non-intrusive metrics to evaluate speech quality due to the challenges of using intrusive metrics. This is because real-world lip-to-speech tasks often involve unconstrained environments with background noise and speaker variability. Intrusive metrics, commonly used in previous lip-to-speech work, aim to perfectly reconstruct the original speech waveform, which is also unnecessary and impractical for speech generation in unconstrained datasets like LRS2 and LRS3.

We compare NaturalL2S with several baselines across various aspects: speech quality, content preservation, audio-visual synchronization, prosody, and speaker similarity. For speech quality, we use non-intrusive metrics STOI-Net\cite{zezario2020stoi} and DNSMOS\cite{reddy2021dnsmos}. In addition, we measure Lip Sync Error - Distance (LSE-D) and Lip Sync Error - Confidence (LSE-C) proposed by \cite{chung2017}, where LSE-D measures the distance between the lip and audio represantions, and LSE-C measures the average confidence of the synchronization. To assess intelligibility, we leverage the powerful Auto-AVSR model [7], trained on LRS2, LRS3, AVSpeech\cite{ephrat2018looking}, and VoxCeleb2, for evaluation. We measure prosodic consistency between generated and target speech using the Pearson Correlation Coefficient (PCC) of normalized F0 sequences, which can be formulated as:
\begin{equation}
  \label{deqn_ex1a} 
  \rho\left(F0, \widehat{F0}\right)=\frac{\operatorname{cov}\left(F0, \widehat{F0}\right)}{\sigma_{F0} \sigma_{\widehat{F0}}},
\end{equation}
where \(\sigma_{F0}\) and \(\sigma_{\widehat{F0}}\) are the standard deviations of \(F0\) and \(\widehat{F0}\), respectively. And the cov (·) represents the covariance between the F0 sequences. A higher PCC value indicates better performance.
Finally, we employ the deep learning package Resemblyzer\cite{wan2018generalized}, a speaker embedding model trained through contrastive learning, to evaluate voice similarity. In addition to these objective metrics, we also conduct human subjective MOS tests to assess the quality, naturalness, and intelligibility of the generated speech samples.

\begin{figure*}[!h] 
	\centering
	\subfloat[]{\includegraphics[width=7in]{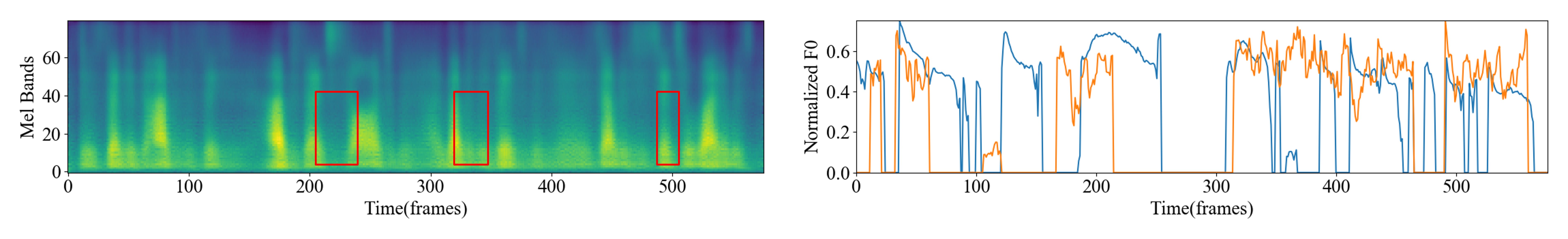}\label{fig3:subfig1}}
	\\* [0.01ex]
	\subfloat[]{\includegraphics[width=7in]{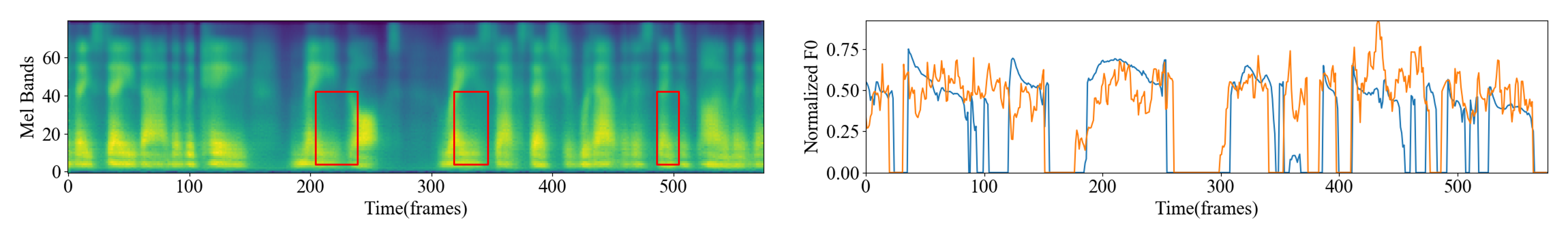}\label{fig3:subfig2}}
	\\* [0.01ex]
	\subfloat[]{\includegraphics[width=7in]{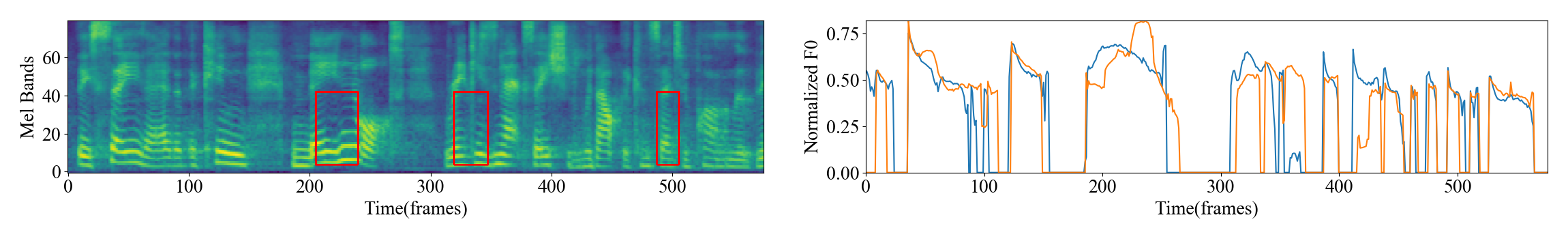}\label{fig3:subfig3}}
	\\* [0.01ex]
	\subfloat[]{\includegraphics[width=7in]{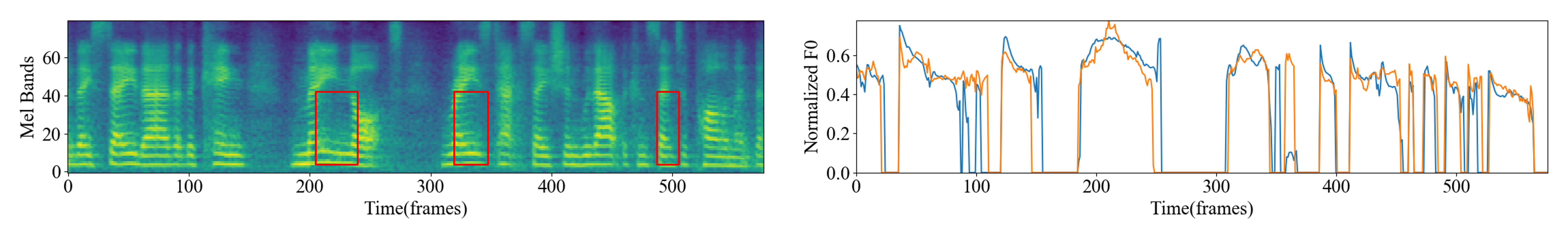}\label{fig3:subfig4}}
	\\* [0.01ex]
	\subfloat[]{\includegraphics[width=7in]{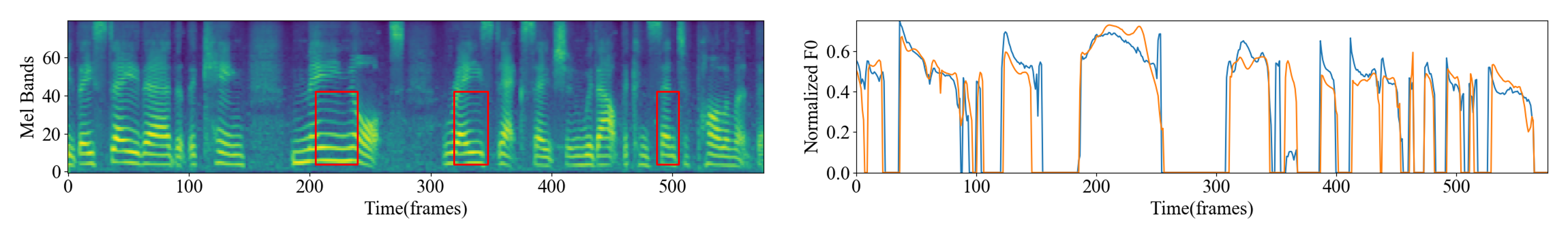}\label{fig3:subfig5}}
	\\* [0.01ex]
	\subfloat[]{\includegraphics[width=7in]{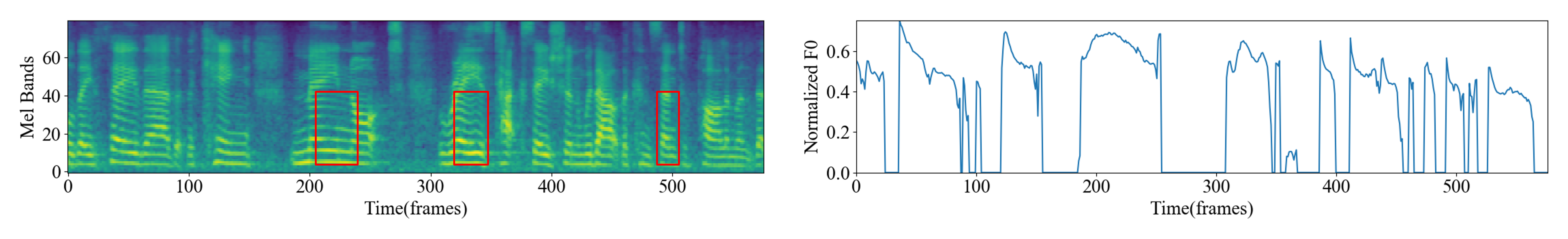}\label{fig3:subfig6}}
	\caption{\centering Qualitative examples on the LRS2 datasets generated by (a) VCA-GAN\cite{kim2021a}, (b) Multi-task\cite{kim2023-liptospeech}, (c) DiffV2S\cite{choi2023}, (d) Lip2Speech Unit\cite{choi2023-intelligible}, (e) NaturalL2S, (f) Ground truth. The left part of each subfigure is the mel-spectrogram, and the right part is the normalized fundamental frequency curve.}
	\label{fig:figure3}
\end{figure*}

\section{Results}
In this section, we present both objective and subjective evaluation results of our comparative experiments to validate the ability of the proposed model to generate high-quality speech. Additionally, ablation studies will be conducted to analyze the effectiveness of crucial modules and training strategies within the model. For the comparative experiments, our baselines include VCA-GAN\cite{kim2021a}, Multi-task\cite{kim2023-liptospeech}, DiffV2S\cite{choi2023} and Lip2Speech Unit\cite{choi2023-intelligible}. The VCA-GAN and Multi-task are trained with the official open-source code. Due to the inference code for the LRS2 dataset in Lip2Speech Unit is missing and the code implementation of DiffV2S is not open-source, we rely on synthesized video samples provided by the authors for comparison.

\subsection{Objective Evaluation Results}
We start by reporting the qualitative performance. Results on LRS2 and LRS3 are summarized in Table \ref{tbl:1} and Table \ref{tbl:2} respectively. The evaluation metrics encompass speech quality, intelligibility, prosody, voice similarity, and audio-visual synchronization. It is worth noting that the pre-trained AV-HuBERT is utilized in DiffV2S, Lip2Speech Unit and our model. 

The objective evaluation results in Tables \ref{tbl:1} and \ref{tbl:2} validate the effectiveness of the proposed NaturalL2S model. As expected, methods utilizing both a pre-trained AV-HuBERT model and a neural vocoder (DiffV2S, Lip2Speech Unit, and NaturalL2S) achieve significantly higher speech quality scores (STOI-Net, DNSMOS) compared to those without one. The pre-trained AV-HuBERT plays a crucial role in generating accurate spectrograms, as evidenced by the improved performance. This also highlights the superiority of neural vocoders in synthesizing natural-sounding speech. However, NaturalL2S still demonstrates a significant improvement in speech quality compared to other baselines using AV-HuBERT (except DiffV2S) due to the introduction of acoustic prior information. It is important to note that NaturalL2S achieves even better speech intelligibility as measured by WER compared to DiffV2S. The pre-trained AV-HuBERT also significantly improves speech intelligibility compared to methods without it, as observed in the WER metric. NaturalL2S outperforms other competitors in WER (excluding Lip2Speech Unit), demonstrating its effectiveness in generating understandable speech. Notably, our model achieves comparable WER scores with the Lip2Speech Unit but surpasses it in other metrics like speech quality and synchronization. Furthermore, the end-to-end training strategy in NaturalL2S leads to superior audio-visual synchronization. NaturalL2S also achieves higher prosody similarity to the target speech compared to other baselines. Finally, our method achieves the best voice similarity on the LRS3 dataset without using a reference speaker embedding and delivers equally optimal performance as DiffV2S on the LRS2 dataset. This suggests that NaturalL2S can learn speaker information without additional supervision, making it more practical in real-world scenarios.

\begin{table}[!h]
\renewcommand{\arraystretch}{1.2}
  \renewcommand{\tabcolsep}{0.7mm}
	\setlength{\abovecaptionskip}{0pt}
	\centering
	\footnotesize
	\caption{Human Mean Opinion Scores and Standard Deviations for Different Models} \label{tbl:3}
	\begin{tabular}{lcccc}
		\toprule
		\textbf{Model}  & Quality & Naturalness & Intelligibility & Voice Similarity\\
		\midrule
		VCA-GAN  & 1.16$\scriptstyle{\pm 0.28}$ & 1.16$\scriptstyle{\pm 0.29}$ & 1.23$\scriptstyle{\pm 0.35}$ & 1.17$\scriptstyle{\pm 0.31}$ \\
		Multi-task  & 1.64$\scriptstyle{\pm 0.25}$ & 1.54$\scriptstyle{\pm 0.33}$ & 2.06$\scriptstyle{\pm 0.61}$ & 1.54$\scriptstyle{\pm 0.37}$ \\
		DiffV2S  & 3.88$\scriptstyle{\pm 0.39}$ & 3.61$\scriptstyle{\pm 0.53}$ & 3.83$\scriptstyle{\pm 0.44}$ & 3.30$\scriptstyle{\pm 0.55}$ \\
		Lip2Speech Unit & 3.29$\scriptstyle{\pm 0.28}$ & 3.34$\scriptstyle{\pm 0.37}$ & 3.71$\scriptstyle{\pm 0.38}$ & 3.27$\scriptstyle{\pm 0.28}$  \\
		\textbf{Proposed} & \textbf{4.40$\scriptstyle{\pm 0.26}$} & \textbf{4.34$\scriptstyle{\pm 0.30}$} & \textbf{4.39$\scriptstyle{\pm 0.32}$} & \textbf{3.98$\scriptstyle{\pm 0.33}$} \\
		Ground Truth & 4.88$\scriptstyle{\pm 0.24}$ & 4.94$\scriptstyle{\pm 0.16}$ & 4.93$\scriptstyle{\pm 0.16}$ & -- \\
		\bottomrule
	\end{tabular}
\end{table}

Fig. \ref{fig:figure3} showcases qualitative examples that further support the effectiveness of NaturalL2S. It includes mel-spectrograms and normalized pitch (F0) for both ground truth and generated speech. As observed, the generated speech of NaturalL2S addresses issues like harmonic distortion and over-smoothing present in other baselines, leading to improved naturalness and quality. Additionally, the generated speech of NaturalL2S may even incorporate richer harmonic signals compared to the ground truth, which potentially explains the higher quality scores achieved by our method. The F0 plots demonstrate that the generated F0 curves by NaturalL2S align the ground truth well with the pitch changes. These qualitative observations support the trends highlighted in the quantitative metrics reported in Tables \ref{tbl:1} and \ref{tbl:2}.

\subsection{Subjective Evaluation Results}
In the lip-to-speech task, subjective evaluation is vital to assess how humans perceive the generated speech. Therefore, we conducted a mean opinion score (MOS) listening test by 15 participants to evaluate real human perception of the generated speech. We randomly selected 50 samples each from the LRS2 and LRS3 test splits for both our NaturalL2S model and the baseline models. Listeners were asked to rate the different evaluative dimensions: quality, naturalness, intelligibility, and voice similarity. Each sample was scored using a 5-point scale, with 5 being excellent, 4 good, 3 fair, 2 poor, and 1 bad. Table \ref{tbl:3} reports the MOS scores and standard deviations of NaturalL2S and other baselines. 

As shown in Table \ref{tbl:3}, the proposed NaturalL2S achieves superior performance on all subjective metrics compared to the competitors. In terms of speech quality, NaturalL2S significantly surpasses all baselines, obtaining a MOS score of 4.40. This substantial improvement suggests that the end-to-end training strategy and the incorporation of acoustic prior information by DDSP effectively enable NaturalL2S to generate high-quality speech. These findings are consistent with the positive results observed in the objective evaluation. Similar trends are observed for naturalness, intelligibility, and voice similarity, demonstrating the overall effectiveness of NaturalL2S. While the NaturalL2S exhibits slightly higher WER scores compared to the Lip2Speech Unit, the MOS results suggest that high-quality speech significantly improves intelligibility for human listeners. This phenomenon is also reflected in the voice similarity MOS scores. These results highlight the importance of generated speech quality and naturalness in lip-to-speech tasks. In conclusion, the comparison results support the advantage of the proposed NaturalL2S in generating high-quality speech.

\subsection{Ablation Study}
To validate the effectiveness of each module in NaturalL2S and its contribution to the performance of the model, we conducted an ablation study on the LRS2 dataset. Specifically, we studied the impact of each module of NaturalL2S on the model performance, including end-to-end training, DDSP module, content supervision and mel-predictor. The results are presented in Table \ref{tbl:4} using the following abbreviations for the models:

\begin{table*}[ht]
	\setlength{\abovecaptionskip}{0pt}
	\centering
	\footnotesize
	\caption{Ablation study. Performance evaluation between different variances of NaturalL2S on LRS2 dataset.} \label{tbl:4}
	\begin{tabular}{lcccccccc}
		\toprule
		\textbf{Model}  & STOI-Net ↑ & DNS-MOS ↑ & WER ↓ & LSE-C ↑ & LSE-D ↓ & F0pcc ↑ & Voice Similarity ↑  \\
		\midrule
		Full NaturalL2S  & \textbf{0.92} & \textbf{2.95} & 36.5\% & 8.74 & 5.84 & \textbf{0.61} & \textbf{0.63} \\  
		NaturalL2S-wo-e2e  & 0.85 & 2.39 & \textbf{36.3\%} & 8.03 & 6.34 & 0.57  & 0.53 \\
		NaturalL2S-wo-DDSP & 0.91 & 2.89 & 37.0\% & 8.52 & 5.94 & 0.60 & 0.63 \\
		NaturalL2S-wo-unit & 0.92 & 2.92 & 38.8\% & 8.60 & 5.88 & 0.60 & 0.62 \\
		NaturalL2S-wo-mel & 0.92 & 2.88 & 36.7\% & \textbf{8.76} & \textbf{5.79} & 0.61 & 0.63 \\
		\bottomrule
	\end{tabular}
\end{table*}

\subsubsection{NaturalL2S-wo-e2e}
NaturalL2S training and inference are not performed in an end-to-end manner. We divide the training process into two stages. In the first stage, we remove the pitch generator and DDSP synthesizer, focusing only on the content generator, which generates mel-spectrograms instead of content embeddings optimized by the reconstruction loss. We then train a vanilla HiFi-GAN vocoder on the LRS3 dataset to generate speech from these mel-spectrograms. During inference, we cascade the two models to generate waveforms.

\subsubsection{NaturalL2S-wo-DDSP}
NaturalL2S without the DDSP synthesizer, which removes the pitch generator and converts the conditional HiFi-GAN to a vanilla HiFi-GAN.

\subsubsection{NaturalL2S-wo-unit}
NaturalL2S without training the FC layer, thus discarding content supervision of speech units.

\subsubsection{NaturalL2S-wo-mel}
NaturalL2S without the use of mel-predictor and F0 post-predictor. The output of the F0 predictor is directly passed through the DDSP synthesizer.

\begin{figure}[htbp] 
	\centering
	\subfloat[]{\includegraphics[width=3.4in]{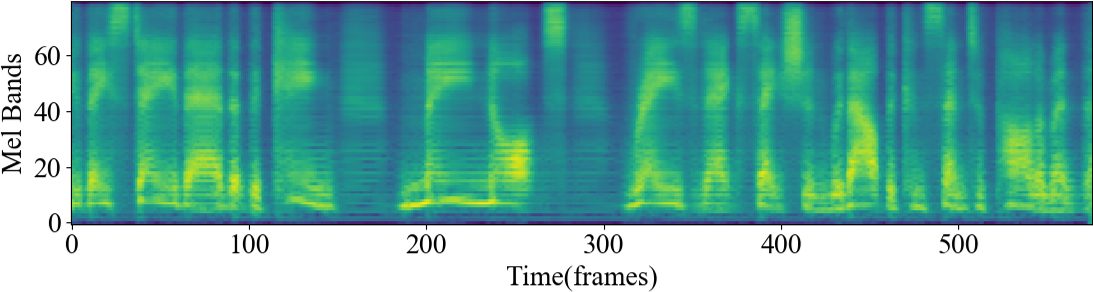}\label{fig4:subfig1}}
	\\* [0.01ex]
	\subfloat[]{\includegraphics[width=3.4in]{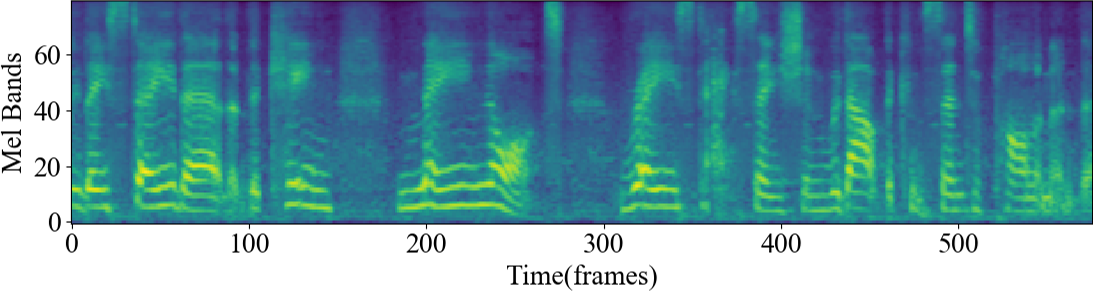}\label{fig4:subfig2}}
	\\* [0.01ex]
	\subfloat[]{\includegraphics[width=3.4in]{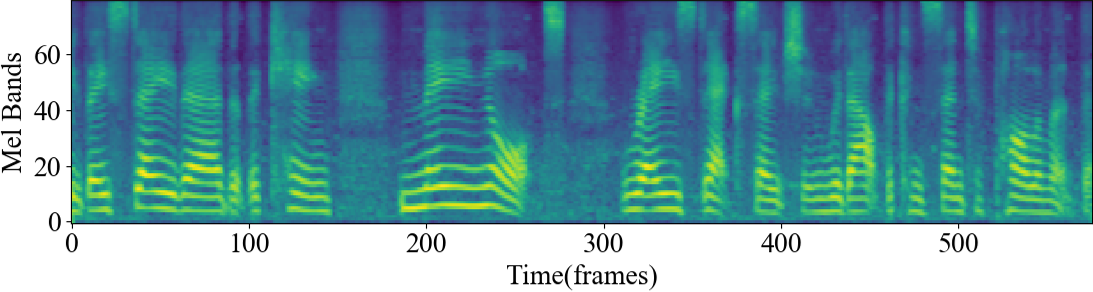}\label{fig4:subfig3}}
	\\* [0.01ex]
	\subfloat[]{\includegraphics[width=3.4in]{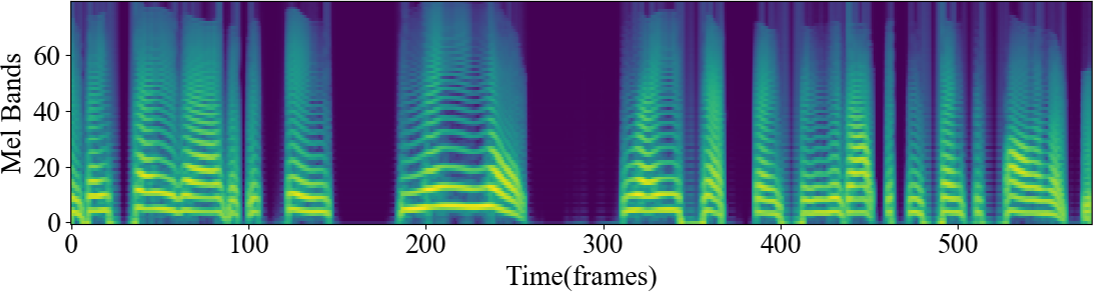}\label{fig4:subfig4}}
	\\* [0.01ex]
	\subfloat[]{\includegraphics[width=3.4in]{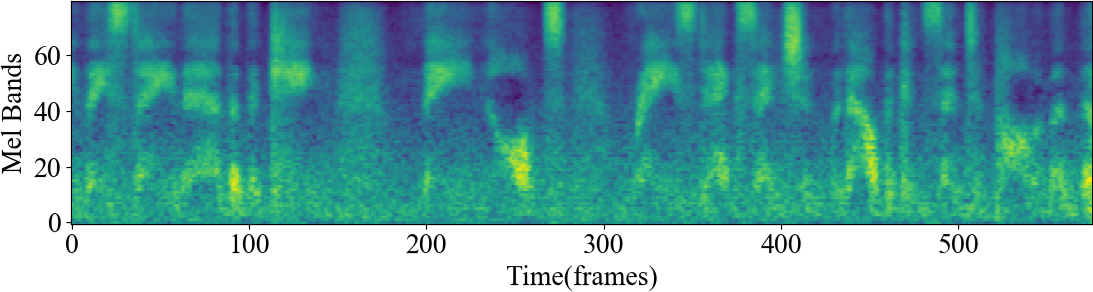}\label{fig4:subfig5}}
	\caption{Qualitative results generated on the ablation study. (a) NaturalL2S-wo-e2e, (b) NaturalL2S-wo-DDSP, (c) Full NaturalL2S, (d) Harmonic signal from harmonic synthesizer, (e) Noise signal from harmonic synthesizer.}
	\label{fig:figure4}	
\end{figure}

Comparing the evaluation results for \emph{Full NaturalL2S} and \emph{NaturalL2S-wo-e2e} from Table \ref{tbl:4}, although \emph{NaturalL2S-wo-e2e} shows a slight improvement in WER, it suffers in speech quality, audio-visual synchronization, prosody, and voice similarity. This highlights the importance of the end-to-end training strategy. By jointly training all components, the model learns efficient hidden representations that capture the complex relationship between visual features and speech. In contrast, directly generating a mel-spectrogram in NaturalL2S-wo-e2e leads to over-smoothing issues, plotted in Fig. \ref{fig4:subfig1} and Fig. \ref{fig4:subfig3}, due to limited information from lip movements and insufficient model capacity. The mismatch between vocoder training and inference further destroys speech quality. Although \cite{choi2023-intelligible} leverages data augmentations to relieve the mismatch problem, the noise is also introduced to the generated speech waveforms by the augmentation technique. We believe that end-to-end train the vocoder is an effective way to route around the aforementioned troubles and generate high-quality speech waveforms. From Table \ref{tbl:4}, we can also observe that the prosody consistency and voice similarity also decrease because of the failure in speech quality and the absence of the pitch generator.

Another important result in Table \ref{tbl:4} can be observed in the comparison between \emph{Full NaturalL2S} and \emph{NaturalL2S-wo-DDSP}. In terms of prosody consistency and voice similarity, there is no marked difference between the two methods. However, the speech quality and audio-visual synchronization are worse than the proposed method. The corresponding mel-spectrogram (Fig. \ref{fig4:subfig2}) lacks the richer harmonic structure introduced by the DDSP synthesizer in the full model. Fig. \ref{fig4:subfig4} and Fig. \ref{fig4:subfig5} visualize the separated harmonic and noise waveforms generated by the DDSP synthesizer. The harmonic signal provides clear pitch information, while the noise component refines the aperiodic parts. The Conditional HiFi-GAN then takes this coarse signal and generates more natural speech. Overall, the DDSP synthesizer plays a crucial role in recovering high-quality speech compared to \emph{NaturalL2S-wo-DDSP}. Additionally, Table \ref{tbl:4} demonstrates that the acoustic inductive bias from the DDSP module improves overall metrics except for voice similarity.

Removing content supervision \emph{NaturalL2S-wo-unit} leads to a higher WER and slightly worse performance on other metrics through Table \ref{tbl:4}. While the AV-HuBERT embeddings already contain semantic information, the additional linguistic supervision from audio HuBERT further improves the final result.

Finally, the performance of \emph{NaturalL2S-wo-mel} reported in Table \ref{tbl:4} is comparable to the full model in most cases except for DNSMOS. This suggests that video features alone can be sufficient for F0 prediction, while the content embedding provides additional benefits for the pitch generator.

The ablation study effectively validates the importance of each module in NaturalL2S.  We demonstrate that the end-to-end training strategy, DDSP synthesizer, content supervision, and mel-predictor all contribute to the model in generating high-quality and natural speech.

\section{Conclusion}
In this paper, we proposed NaturalL2S, an end-to-end lip-to-speech system that generates high-quality speech with high intelligibility from silent videos. Our system incorporates the differentiable digital signal processing synthesizer that leverages acoustic prior information to generate speech waveform. Evaluations on the "in-the-wild" multi-speaker datasets demonstrate that NaturalL2S achieves state-of-the-art performance in terms of speech quality and audio-visual synchronization. While the pre-trained AV-HuBERT plays a crucial role in feature extraction for our system, the effectiveness of NaturalL2S lies in combined different technologies, particularly the end-to-end training strategy and the differentiable digital signal processing synthesizer.

Although promising performance has been achieved recently, there exist many critical issues for further investigation. Exploring lip-to-speech methods without relying on large pre-trained models by incorporating acoustic priors is a promising direction. What is more, developing lightweight and practical methods is also crucial for resource-limited application scenarios.

\end{document}